\documentclass[aps,twocolumn,amsmath,amssymb,nofootinbib,superscriptaddress]{revtex4-2}
\usepackage{adjustbox}
\usepackage[utf8]{inputenc} 
\usepackage[colorlinks=true,
					linkcolor=magenta,
					anchorcolor=magenta,
					citecolor=magenta,
					urlcolor=magenta]{hyperref}        
\usepackage{url}            
\usepackage{booktabs}       
\usepackage{amsfonts}       
\usepackage{nicefrac}       
\usepackage{microtype}      
\usepackage{natbib}

\usepackage{bm}
\usepackage{bbm}
\usepackage{braket}
\usepackage{graphicx}
\usepackage{subfig}
\usepackage{float}
\usepackage{amsthm}
\usepackage{algorithmic}
\usepackage[linesnumbered,ruled]{algorithm2e}
\SetKwInOut{Parameter}{parameter}

\usepackage{mathtools}
\usepackage{nccmath}
\usepackage{color}
\usepackage{caption}
\usepackage{titlesec}

\usepackage{soul}
\titlespacing{\title}{0pc}{0.1pc}{0.3pc}
\titlespacing{\section}{0pc}{0.1pc}{0.3pc}

\begin{document}

\title{Universality of universal single-qubit-gate decomposition with coherent errors}

\author{Ruixia Wang}
\email{wangrx@baqis.ac.cn}
\affiliation{Beijing Academy of Quantum Information Sciences, Beijing 100193, China}%

\author{Peng Zhao}
\email{shangniguo@sina.com}
\affiliation{Beijing Academy of Quantum Information Sciences, Beijing 100193, China}%

\author{Haifeng Yu}
\email{hfyu@baqis.ac.cn}
\affiliation{Beijing Academy of Quantum Information Sciences, Beijing 100193, China}%

\date{\today}

\begin{abstract}
To generate arbitrary one- and two-qubit gates, the universal decompositions are usually used in quantum computing, and the universality of these decompositions has been demonstrated. However, in realistic experiments, gate errors may affect the universality of the universal decompositions. Here, we focus on the single-qubit-gate decomposition scheme and study the coherent-error effects on universality. We prove that, in the parameter space which we studied, some kinds of coherent errors will not affect the original universality, but some others will destroy it.
We provide the definition and analytical solutions for universality with coherent errors and propose methods to resume the accuracy of the operations with coherent errors based on our analysis. We also give the analytical results for three kinds of fidelities, which provide another metric for universality and comprehensively depict the resilience of the decomposition scheme with various kinds of coherent errors. Our work introduces a different way of thinking for quantum compilation than existing methods.
\end{abstract}

\maketitle

\preprint{AIP/123-QED}

\emph{Introduction}--Universality is an issue of crucial importance in physical implementations of quantum information processing \cite{zhang2003geometric,nielsen2002quantum}. It is known that quantum computation with one- and two-qubit gates is universal and arbitrary desired quantum operations can be decomposed into a sequence of these base gates \cite{barenco1995elementary}. The universal gate libraries \cite{barenco1995elementary,divincenzo1995two}, minimization of gate counts \cite{harrow2002efficient,vidal2004universal,shende2004minimal} and optimal universal gate decomposition schemes \cite{shende2004minimal,mckay2017efficient,kliuchnikov2013asymptotically,earnest2021pulse} have been studied extensively. Based on the assumption that the base gates for quantum-gate decomposition can be implemented ideally, the theoretical derivations and proofs of universal decomposition schemes have been proposed \cite{mckay2017efficient,shende2004minimal}, and theoretically, these schemes have the highest expressivity \cite{sim2019expressibility,du2020expressive,holmes2022connecting,du2022efficient} for the quantum circuits with corresponding qubit numbers. However, in realistic experiments, on account of the unavoidable errors, e.g. pulse distortion \cite{hincks2015controlling,rol2020time} and microwave crosstalk \cite{arute2019quantum,sarovar2020detecting,zhao2022quantum}, the base gates used in the decomposition schemes cannot be implemented ideally with unit fidelities \cite{kwon2021gate}.

In all kinds of experimental errors, the coherent error is ubiquitous in many quantum systems, and  in some instances, it is more damaging than the incoherent errors, e.g. in the context of quantum error correction \cite{cai2020mitigating}. Coherent error is unitary and will cause, such as, an unwanted rotation of the single qubit, and limit the performance of the quantum circuits \cite{Hashim2021Randomized,krinner2020benchmarking,ouyang2021avoiding,yang2022effect}. So that, there are many schemes which are designed to suppress the coherent errors \cite{zhang2022hidden,bennett1996purification,emerson2007symmetrized,debroy2018stabilizer}, and hoping to eventually correct them by quantum error corrections \cite{debroy2018stabilizer,beale2018quantum,huang2019performance,iverson2020coherence,bravyi2018correcting}. There are also some proposed error mitigation methods, which leverage the optimizable parameters in universal one- and two-qubit-gate decomposition schemes and utilize the optimization approach to correct gate errors and improve the fidelities of the intended operations \cite{maldonado2022error,lao2022software,wang2022control}. However, how the coherent errors affect and when they will disrupt the universality of the universal quantum-gate decomposition schemes have not been systematically studied. Understanding such mechanisms will be helpful for us to suppress or even totally corrected the coherent errors and it can enlighten the thoughts of quantum compilation as well.

In this work, we focus on one of the frequently used universal single-qubit-gate decomposition schemes, study the coherent-error effects on universality, and answer the following questions. 1. Referring to the universality in ideal conditions as the original universality, which kind of coherent errors will destroy the original universality? 2. How to express universality with coherent errors quantitatively? 3. Referring to the gate fidelity with coherent errors and without (with) error mitigation (error mitigation specifically refers to the optimization of the parameters in the decomposition scheme) as original (best) fidelity, what are the analytical expressions for original and best gate fidelities when using the universal decomposition scheme  and what do they reveal? We show that, depending on our classification of the coherent errors in the parameter space, there is one kind of coherent error that will destroy the original universality, while others will not. We give the definition of universality, as well as the analytical expressions of it with various kinds of coherent errors, which elaborate the relationship between the universality and coherent errors. Furthermore, the analytical solutions for the original and best fidelities are given. We show that there are close relationships between coherent errors, the best fidelity and universality for the single-qubit-gate decomposition scheme. 

\begin{figure}[t!]
\centering{\includegraphics[width=80mm]{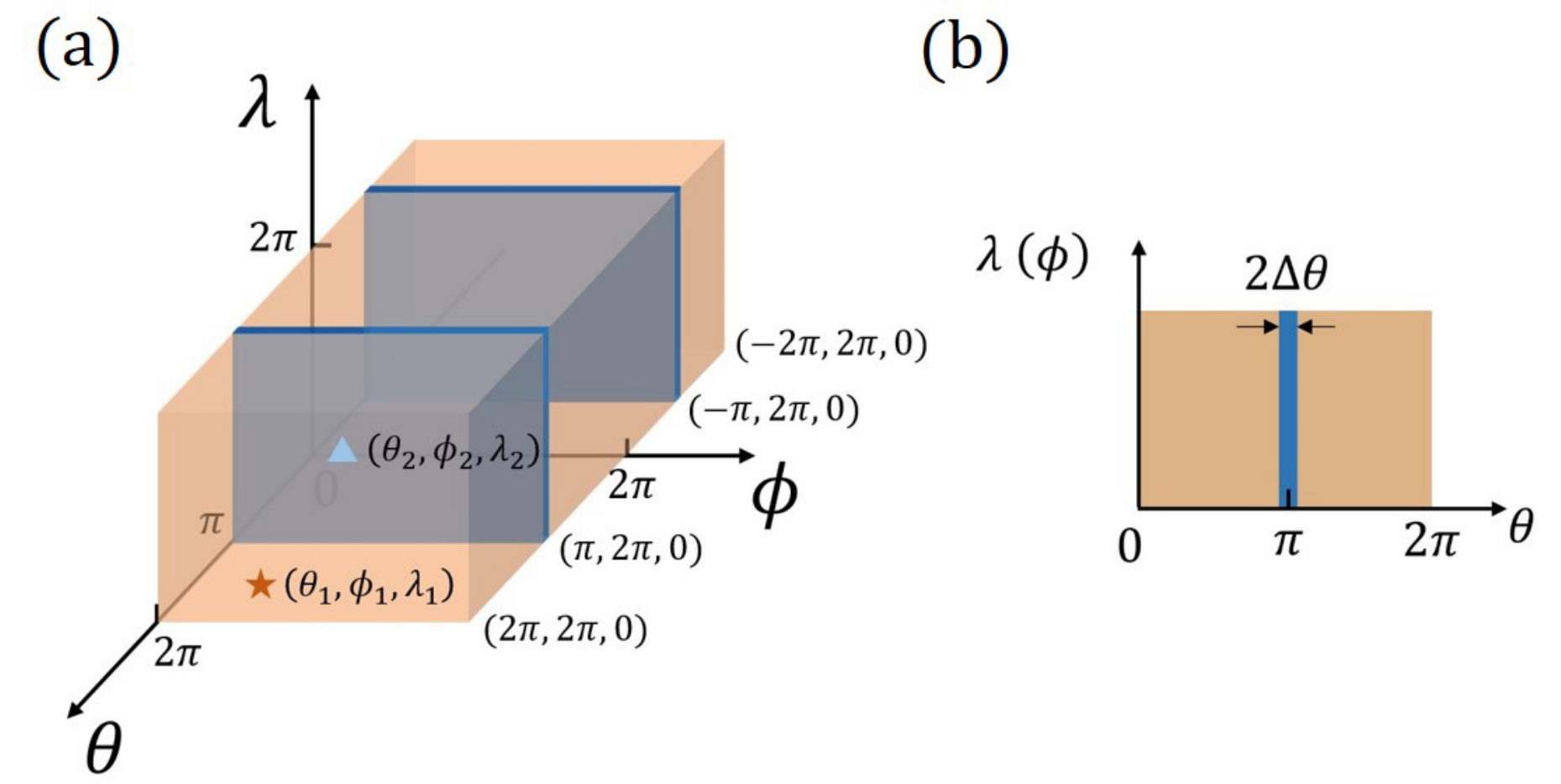}}
\caption{(a) The parameter space for $\{\theta,\phi,\lambda\}$ in three-dimensional Cartesian coordinate system. In the main text, removing the symmetries, we restrict the parameter space which we studied to $\theta\in[0,\pi]$, $\phi\in[0,2\pi)$ and $\lambda\in[0,2\pi)$. (b) Cross sections for the $\theta\lambda$ and $\theta\phi$ planes. The blue areas for (a) and (b) represent the parameter space which cannot be covered by the decomposition scheme with coherent errors.} 
\label{f1}
\end{figure}

\emph{Universal decomposition}--The universal single-qubit gate can be represented by a unitary matrix with three parameters as \cite{mckay2017efficient}

\begin{eqnarray}
U(\theta,\phi,\lambda)=\left[\begin{array}{cc}
\cos(\theta/2) & -ie^{i\lambda}\sin(\theta/2)\\
-ie^{i\phi}\sin(\theta/2) & e^{i(\lambda+\phi)}\cos(\theta/2)
\end{array}\right],
\label{eqM}
\end{eqnarray}
where the parameter $\theta$ has a period of $4\pi$, $\phi$ and $\lambda$ have periods of $2\pi$. We use the three-dimensional Cartesian coordinate system to represent the parameter space, as shown in figure \ref{f1}(a). Because there is $U(\theta+2\pi,\phi,\lambda)=e^{i\pi}U(\theta,\phi,\lambda)$, the matrices with $\theta$ and $\theta+2\pi$ differ by only one global phase, which are equivalent in quantum operations, then we can narrow the studied parameter space of $\theta$ to a period of $2\pi$. Moreover, there is $U(-\theta,\phi,\lambda)=U(\theta,\phi+\pi,\lambda+\pi)$, removing the symmetries, we can restrict the parameter space to $\theta\in[0,\pi]$, $\phi\in[0,2\pi)$ and $\lambda\in[0,2\pi)$. For convenience, in the following text, we use $U$ to represent $U(\theta,\phi,\lambda)$ when necessary.

For most of the quantum computing systems, the Z rotations can be realized by adding a phase offset to the driving field for all subsequent X and Y rotations, which is known as a virtual Z gate, which is essentially perfect \cite{mckay2017efficient,johnson2015demonstration}. By applying the advantage of the virtual Z gate, arbitrary single qubit operation shown in equation (\ref{eqM}) can be composed by a five-gate sequence \cite{mckay2017efficient}:

\begin{equation}
U=Z_{\phi-\pi/2}X_{\pi/2}Z_{\pi-\theta}X_{\pi/2}Z_{\lambda-\pi/2}.
\label{eqUd}
\end{equation}
If the Z and X rotation gates in equation (\ref{eqUd}) are ideal with unit fidelities, this decomposition is universal, and equations (\ref{eqM}) and (\ref{eqUd}) are equivalent. When we apply equation (\ref{eqUd}) in realistic quantum device, the errors with virtual Z rotations can be neglected, but the $X_{\pi/2}$ gate is operated by applying the driving pulse, the errors with which cannot be neglected. Depending on equation (\ref{eqM}), the $X_{\pi/2}$ gate can be expressed as $U_{X_{\pi/2}}=U(\theta_x,\phi_x,\lambda_x)$, and the parameters take the values of $\{\theta_x, \phi_x, \lambda_x\} = \{\pi/2, 0, 0\}$ ideally. However, when coherent errors happen, at least one of the following cases occur, which are $\theta_x=\pi/2+\delta$ ($\delta\ne 0$), $\lambda_x\ne 0$ and $\phi_x\ne 0$. Substituting $U_{X_{\pi/2}}$ into equation (\ref{eqUd}) and replacing the operations of $X_{\pi/2}$, extracting and discarding the global phase, we can get \cite{SM} $\tilde{U}=\left[\tilde{U}^{(11)}, \tilde{U}^{(12)}; \tilde{U}^{(21)}, \tilde{U}^{(22)}\right]$,
with the definition of $a_{+}\equiv\lambda_x+\phi_x$, there are

\begin{eqnarray}
\tilde{U}^{(11)}&=&e^{i\theta/2}\cos^{2}(\theta_{x}/2)+e^{i(-\theta/2+a_{+})}\sin^{2}(\theta_{x}/2),\\
\tilde{U}^{(12)}&=&[e^{i(-\theta/2+\lambda+\lambda_{x}+a_{+})}-e^{i(\theta/2+\lambda+\lambda_{x})}]\frac{\sin\theta_{x}}{2},\\
\tilde{U}^{(21)}&=&[e^{i(-\theta/2+\phi+\phi_{x}+a_{+})}-e^{i(\theta/2+\phi+\phi_{x})}]\frac{\sin\theta_{x}}{2},\\
\tilde{U}^{(22)}&=&e^{i(\theta/2+\lambda+\phi+a_{+})}\sin^{2}(\theta_{x}/2)\nonumber\\
&&+e^{i(-\theta/2+\lambda+\phi+2a_{+})}\cos^{2}(\theta_{x}/2).
\end{eqnarray}

\emph{Universality}--Before talking about universality, we would like to discuss the coverage of $\tilde{U}$ to $U$ in the parameter space in different cases. We divide the coherent errors into three kinds by the parameters on which the coherent error happen. Next, we will discuss the cases with different kinds of coherent errors or combinations of them.

\emph{Case 1}. $\{\theta_x, \phi_x, \lambda_x\}=\{\pi/2+\delta,0,0\}$ ($\delta\ne0$), the coherent error happens on $\theta_x$. Substituting the values of  $\{\theta_x, \phi_x, \lambda_x\}$ into $\tilde{U}$, we can get $\tilde{U}_{Case1}=e^{i\gamma_1}\left[\tilde{U}^{(11)}_{Case1}, \tilde{U}^{(12)}_{Case1}; \tilde{U}^{(21)}_{Case1}, \tilde{U}^{(22)}_{Case1}\right]$ \cite{SM},
where $\tan\gamma_1=\tan(\theta/2)\cos\theta_{x}$, 
\begin{eqnarray}
\tilde{U}^{(11)}_{Case1}&=&\sqrt{1-\sin^{2}(\theta/2)\sin^{2}\theta_{x}},\\
\tilde{U}^{(12)}_{Case1}&=&-ie^{i\lambda_{c1}}\sin(\theta/2)\sin\theta_{x},\\
\tilde{U}^{(21)}_{Case1}&=&-ie^{i\phi_{c1}}\sin(\theta/2)\sin\theta_{x},\\
\tilde{U}^{(22)}_{Case1}&=&e^{i(\lambda_{c1}+\phi_{c1})}\tilde{U}^{(11)}_{Case1},
\end{eqnarray}
and
$\lambda_{c1}\equiv\lambda-\gamma_1$, $\phi_{c1}\equiv\phi-\gamma_1$. Removing the symmetries, we can restrict the parameter space to $\theta_x\in[0,\pi/2) \cup (\pi/2,\pi]$.

Compared with equation (\ref{eqM}), the unitary matrix of $\tilde{U}_{Case1}$ also satisfies the following conditions, $|\tilde{U}^{(11)}_{Case1}|=|\tilde{U}^{(22)}_{Case1}|$, $|\tilde{U}^{(12)}_{Case1}|=|\tilde{U}^{(21)}_{Case1}|$ and $|\tilde{U}^{(11)}_{Case1}|^2+|\tilde{U}^{(12)}_{Case1}|^2=1$. For the item of $\tilde{U}^{(11)}_{Case1}$, the value range is $[|\sin\delta|,1]$. If we define a new parameter $\tilde{\theta}$, which satisfies $\cos(\tilde{\theta}/2)=\ensuremath{\sqrt{1-\sin^{2}(\theta/2)\sin^{2}\theta_{x}}}$ ($\theta_x=\pi/2+\delta$), then within the range of $[0,\pi]$, the value range of $\tilde{\theta}$ is $[0,\pi-2|\delta|]$. Because we have assumed that $\delta\ne0$, then in the parameter space, the value range of $\theta\in(\pi-2|\delta|,\pi]$ cannot be covered. In figure \ref{f1}(a) and (b), we depict the schematic drawing of the area which cannot be covered by $\tilde{U}$ (Here the subscript $Case1$ is omitted, because this colored area can also represent the similar space for other cases, e.g. the $Case3$.) in blue and $\Delta\theta$ represents the width of the uncovered area along the $\theta$ axis. For \emph{Case 1}, there is $\Delta\theta_{Case1}=2|\delta|$. While for the other two parameters, there are $\lambda_{c1}\in[-\gamma_1,-\gamma_1+2\pi)$ and $\phi_{c1} \in[-\gamma_1,-\gamma_1+2\pi)$ with the periods of $2\pi$, then through the translations, they can cover the whole value ranges of $[0,2\pi)$.

\emph{Case 2}. $\theta_x=\pi/2$, $\phi_{x}\ne0$ and $\lambda_{x}\ne0$. In this case, there are coherent errors on both $\phi_{x}$ and $\lambda_{x}$. The operation $U_{X_{\pi/2}}$ can be written as $U^{case2}_{X_{\pi/2}}=Z_{\phi_{x}-\pi/2}X_{\pi/2}Z_{\pi/2}X_{\pi/2}Z_{\lambda_{x}-\pi/2}$. Note that, in the realistic experiment, the operation of $U_{X_{\pi/2}}$ is implemented as a whole and does not be decomposed into several rotations and accomplished separately. But here, we are studying on the coverage of $\tilde{U}$ to $U$, and can assume that, $U_{X_{\pi/2}}$ is decomposed according to equation (\ref{eqUd}). Then, substituting $U^{case2}_{X_{\pi/2}}$ into equation (\ref{eqUd}), we can get $\tilde{U}_{Case2}=Z_{\phi-\pi/2}U^{case2}_{X_{\pi/2}}Z_{\pi-\theta}U^{case2}_{X_{\pi/2}}Z_{\lambda-\pi/2}$. The errors with $U^{case2}_{X_{\pi/2}}$ happen on $\lambda_x$ and $\phi_x$, the corresponding operations of which are located at both ends, so that, it's easy to correct the errors by $Z_{\phi-\pi/2}$, $Z_{\pi-\theta}$ and $Z_{\lambda-\pi/2}$ operations in $\tilde{U}_{Case2}$. Just define $\phi_{c2}\equiv\phi+\phi_{x}$, $\theta_{c2}\equiv\theta-\lambda_{x}-\phi_{x}$, $\lambda_{c2}\equiv\lambda+\lambda_{x}$, then we have $\tilde{U}_{Case2}=Z_{\phi_{c2}-\pi/2}X_{\pi/2}Z_{\pi-\theta_{c2}}X_{\pi/2}Z_{\lambda_{c2}-\pi/2}$. Through the translation operations, the value range of $\{\theta_{c2},\phi_{c2},\lambda_{c2}\}$ can cover which in equation (\ref{eqM}), then $\tilde{U}_{Case2}$ maintains the original universality. In the Supplementary material \cite{SM}, we provide another proof based on the matrix representation.

\emph{Case 3}. $\theta_x=\pi/2+\delta$ ($\delta \ne 0$), $\phi_{x}\ne0$ and $\lambda_{x}\ne0$. In this case, there are coherent errors on all the parameters of $\{\theta_{x}, \phi_{x}, \lambda_{x}\}$. There is $\tilde{U}_{Case3}  =  e^{i\gamma_{3}}\left[\tilde{U}_{Case3}^{(11)}, \tilde{U}_{Case3}^{(12)}; \tilde{U}_{Case3}^{(21)}, \tilde{U}_{Case3}^{(22)}\right]$ \cite{SM},
with the definition of $a_{+}\equiv\lambda_x+\phi_x$, there are

\begin{eqnarray}
\tilde{U}_{Case3}^{(11)} & = & \sqrt{1-\sin^{2}\theta_{x}\sin^2(\frac{\theta-a_{+}}{2})},\\
\tilde{U}_{Case3}^{(12)} & = &  -ie^{i\lambda_{c3}}\sin\theta_{x}\sin|\frac{\theta-a_{+}}{2}|,\\
\tilde{U}_{Case3}^{(21)} & = & -ie^{i\phi_{c3}}\sin\theta_{x}\sin|\frac{\theta-a_{+}}{2}|,\\
\tilde{U}_{Case3}^{(22)} & = & e^{i(\lambda_{c3}+\phi_{c3})}\tilde{U}_{Case3}^{(11)}.
\end{eqnarray}
The specific expressions for $\gamma_3$, $\lambda_{c3}$ and $\phi_{c3}$ can be found in the Supplementary material \cite{SM}. Removing the symmetries, we can restrict the parameter space to $\theta_x\in[0,\pi/2) \cup (\pi/2,\pi]$, $\phi_x\in(0,2\pi)$ and $\lambda_x\in (0,2\pi)$. $\lambda_{c3}$ and $\phi_{c3}$ can cover the whole value ranges of $\lambda$ and $\phi$ in equation (\ref{eqM}) through the coordinate translation operations. We can see that the unitary matrix $\tilde{U}_{Case3}$ satisfies the conditions $|\tilde{U}^{(11)}_{Case3}|=|\tilde{U}^{(22)}_{Case3}|$, $|\tilde{U}^{(12)}_{Case3}|=|\tilde{U}^{(21)}_{Case3}|$ and $|\tilde{U}^{(11)}_{Case3}|^2+|\tilde{U}^{(12)}_{Case3}|^2=1$. The value range of $\tilde{U}^{(11)}_{Case3}$ is also $[|\sin\delta|,1]$, which is the same as \emph{Case 1}, but there is a translation of $a_+$ for the specific value. In the parameter space, the value range of $\theta\in(\pi-2|\delta|,\pi]$ cannot be covered, which is shown in figure \ref{f1}(b) with blue color and there is $\Delta\theta_{Case3}=2|\delta|$. 

Comparing the cases from \emph{Case 1} to \emph{Case 3}, we conclude that, the coherent errors on $\phi_x$ and $\lambda_x$ will not affect the original universality regardless of what value $\theta_x$ takes, while the coherent error on $\theta_x$ will definitively affect the original universality. The coverage of $\tilde{U}$ to $U$ in the parameter space depends on the value of $|\delta|$.
Then according to the above analysis, we give the definition of universality for the universal decomposition with coherent errors. The universality can be defined as

\begin{equation}
UN=\frac{V}{V_{all}},
\end{equation}
where $V_{all}$ is the volume which can be covered by the ideal decomposition in the parameter space of $\{\theta, \phi, \lambda\}$, and $V$ is the volume which can be covered by the decomposition with coherent errors. For the value ranges of $\theta\in[0,\pi]$, $\phi\in[0,2\pi)$ and $\lambda\in[0,2\pi)$, there is $V_{all}=4\pi^{3}$.
For both \emph{Case 1} and \emph{Case 3}, the covered volume is $4\pi^{3}-4\pi^{2}\times2|\delta|$, the universality is $UN_{case1(case3)}=1-\frac{2|\delta|}{\pi}$. For \emph{Case 2}, the universality is $UN_{case2}=1$.

\emph{Fidelities}--The fidelity of quantum operations can be calculated as 
$F  =  \frac{Tr(U_{imp}^{\dagger}U_{imp})+|Tr(U^{\dagger}_{tar}U_{imp})|^{2}}{d(d+1)}$ \cite{pedersen2007fidelity}, $U_{imp}$ is the matrix representation for the implemented operation and $U_{tar}$ is which for the target operation, $d$ is the dimension of the matrix, which equals $2$ for single-qubit gate. 
When we consider an arbitrary single-qubit gate $U_{tar}$ with the decomposition of equation (\ref{eqUd}),  if there are coherent errors with $X_{\pi/2}$ (other kinds of errors, such as the incoherent errors, are not considered in this work), there will be a displacement between $U_{imp}$ and $U_{tar}$ in the parameter space, as well as a reduction with the implemented fidelity. Then the first kind of fidelities that we want to study is the original fidelity, which is the implemented fidelity with coherent errors and without any error mitigations. We denote it as $F_{ori}$.
For this fidelity, the values of the parameters $\{\theta,\phi,\lambda\}$ for the target single-qubit gate are the same as those for the implemented ones denoted as $\{\theta_{imp},\phi_{imp},\lambda_{imp}\}$, i.e., $\{\theta,\phi,\lambda\}=\{\theta_{imp},\phi_{imp},\lambda_{imp}\}$. For simplicity, we unify both notations with $\{\theta,\phi,\lambda\}$.
 
Consider a general case, if there are errors with all three parameters, which means $\theta_x=\pi/2+\delta$ ($\delta\ne0$), $\phi_x\ne0$ and $\lambda_x\ne0$, the original fidelity can be represented as $F_{ori}=\frac{1+2|f_{x}|^2}{3}$ \cite{SM}, where

\begin{eqnarray}
f_{x}&=&\cos(\theta/2)\cos(\theta/2-a_{+}/2)\cos(a_{+}/2)\nonumber\\
&&+\cos(\theta/2)\sin(\theta/2-a_{+}/2)\sin(a_{+}/2)\cos\theta_{x}\nonumber\\
&& +\sin(\theta/2)\sin(\theta/2-a_{+}/2)\cos(a_{-}/2)\sin\theta_{x},
\end{eqnarray}
with $a_{+}\equiv\lambda_x+\phi_x$ and $a_{-}\equiv\lambda_x-\phi_x$. We give some simple solutions for specific cases.

\begin{eqnarray}
F_{ori}&=&1\nonumber\\
 &&{\rm for} \, \{\theta_{x}, \phi_{x}, \lambda_{x}\}=\{\pi/2, 0, 0\},\\
F_{ori}&=&\frac{1+2\cos^{4}(\lambda_{x}/2)}{3} \nonumber\\
 &&{\rm for}\,\{\theta_{x}, \phi_{x}\}=\{\pi/2, 0\}\, {\rm and}\, \lambda_x\ne 0,\\
F_{ori}&=&\frac{1+2\cos^{4}(\phi_{x}/2)}{3} \nonumber\\
 &&{\rm for}\,\{\theta_{x}, \lambda_{x}\}=\{\pi/2, 0\}\, {\rm and}\, \phi_{x}\ne 0,\\
F_{ori}&=&\frac{1+2|f_y|^2}{3}\nonumber\\
&&{\rm for}\,\theta_{x}=\pi/2, \phi_{x}\ne0, \lambda_{x}\ne0,\\
F_{ori}&=&\frac{1+2[1-2\sin^{2}(\delta/2)\sin^{2}(\theta/2)]^{2}}{3}\nonumber\\
&&{\rm for}\, \{\theta_{x}, \phi_{x}, \lambda_{x}\}=\{\pi/2+\delta, 0, 0\},\label{eq5}
\end{eqnarray}
with 
\begin{eqnarray}
f_y&=&\cos\frac{\lambda_{x}+\phi_{x}}{2}\cos\frac{\lambda_{x}}{2}\cos\frac{\phi_{x}}{2}\nonumber\\
&&-\cos(\theta-\frac{\lambda_x+\phi_x}{2})\sin\frac{\lambda_x}{2}\sin\frac{\phi_x}{2}.
\end{eqnarray}
For equation (\ref{eq5}), when $\delta\to 0$, $F_{ori}$ reaches the maximum of unit, when $\sin(\theta/2)=\frac{1}{\sqrt{2}\sin(\delta/2)}$, $F_{ori}$ reaches the minimum value of $\frac{1}{3}$.

The second kind of fidelity is the optimized fidelity, which means we can tune the implemented parameters of $\{\theta_{imp},\phi_{imp},\lambda_{imp}\}$ to correct the coherent errors, we name it as best fidelity and denote it by $F_{best}$.
To obtain the best fidelity for a target single-qubit gate with the parameters of $\{\theta,\phi,\lambda\}$, we can optimize the implemented parameters, and operate the gate with the optimized results of $\{\theta_{imp},\phi_{imp},\lambda_{imp}\}$ to enhance the implemented fidelity. In equation (\ref{eqUd}), the parameters $\{\theta,\phi,\lambda\}$ are only related to the virtual Z gates, so that, the optimizations for these parameters are easy to accomplish in the experiment \cite{mckay2017efficient}. If the intended single-qubit gate resides in the area where the decomposition scheme with coherent errors can cover, i.e., the orange area shown in figure \ref{f1}, such as point $(\theta_1,\phi_1,\lambda_1)$, we can still get a unit fidelity after the optimization for the implemented parameters, i.e., $F_{best}=1$. For example, for $\tilde{U}_{Case3}$, we can tune the values of $\{\theta_{imp},\phi_{imp},\lambda_{imp}\}$ to make $\phi_{c3}=\phi$, $\lambda_{c3}=\lambda$, and $\tilde{U}^{(11)}_{Case3}=\cos(\theta/2)$, to correct the coherent errors and recover the implemented fidelity to the unit. If the intended single-qubit gate resides out of the orange area in figure \ref{f1}, such as point $(\theta_2,\phi_2,\lambda_2)$ in the blue area, the fidelity cannot be recovered to the unit. However, we still can find the nearest point in the orange area to the target one to obtain the highest fidelity.
The best fidelity can be represented as $F_{best}= \frac{1+2\sin^{2}(\theta/2+|\delta|)}{3}$ \cite{SM}. Note that, this formula is for $\theta\in[0,\pi]$, for $\theta\in[-\pi, 0)$, there should be a minus sign before $\theta$ in $F_{best}$. The above theoretical solutions are verified by the numerical simulations \cite{SM}.

\begin{figure}[h!]
\centering{\includegraphics[width=80mm]{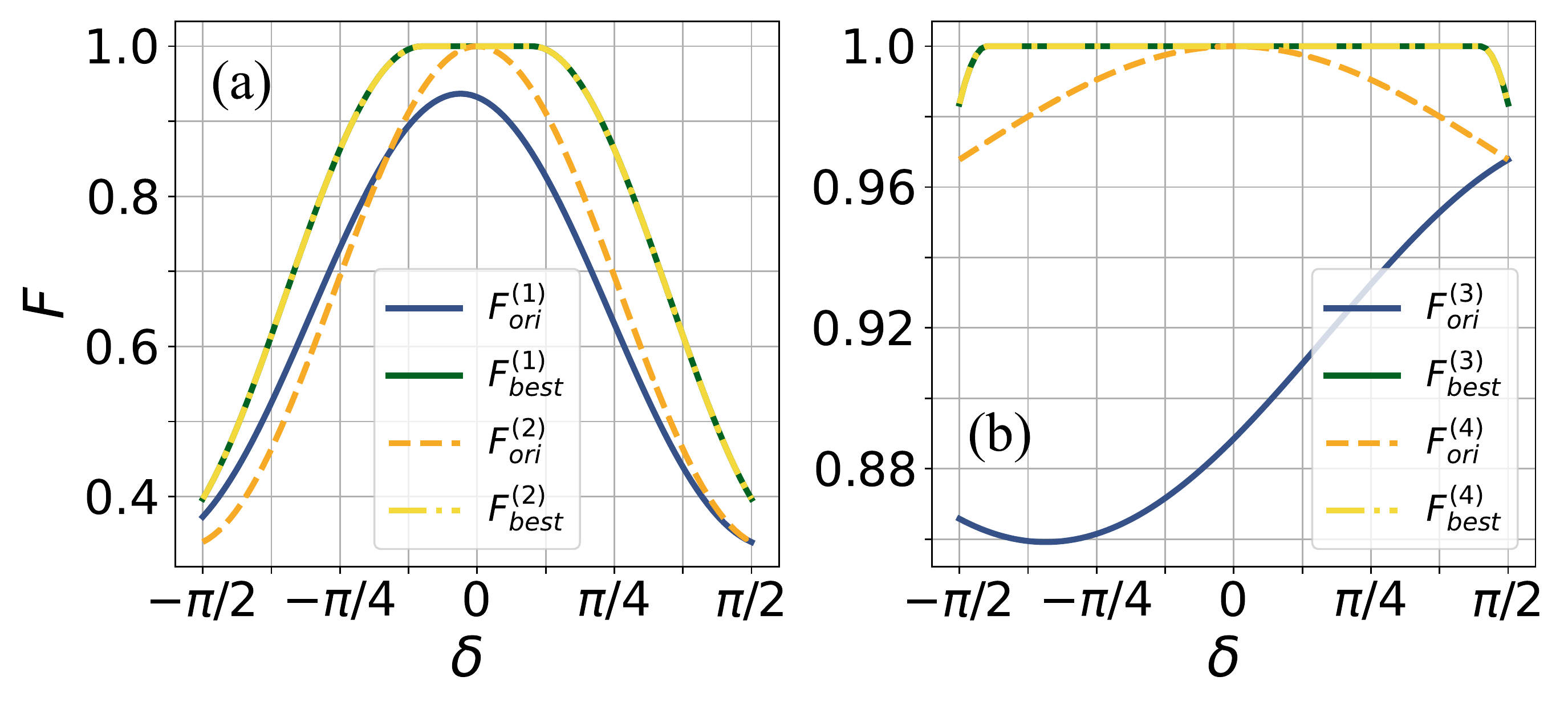}}
\caption{The original fidelity and best fidelity as the value of $\delta$ changes. The parameter values for each example are (a) $\{\theta, \phi, \lambda\}=\{0.8\pi, 1.1\pi, 1.6\pi\}$. (b) $\{\theta, \phi, \lambda\}=\{0.1\pi, 1.1\pi, 1.6\pi\}$. The value of $\theta$ in (a) is near $\pi$, while which in (b) is near $0$, the results show that, as the value of $\theta_x$ aproaches $0$, the decomposition scheme owns greater resilience with coherent errors. For $F^{(1)}_{j}$ in (a) and $F^{(3)}_{j}$ in (b), there are $\{\phi_x, \lambda_x\}=\{0.1\pi, 0.09\pi\}$, and for $F^{(2)}_{j}$ in (a) and $F^{(4)}_{j}$ in (b), there are $\{\phi_x, \lambda_x\}=\{0, 0\}$, $j$ can be $ori$ or $best$. The results show that, the best fidelity depends only on the value of $\theta_x$. The asymmetry of the original fidelity with the axis of $\delta=0$ is induced by $\phi_x$ and $\lambda_x$.}
\label{f2}
\end{figure}

In figure \ref{f2}, we show the information which the fidelities reveal. First, as long as there are coherent errors on any one of the parameters $\{\theta_x,\phi_x,\lambda_x\}$, the best fidelity will always be larger than the original one. Although the error on $\theta_x$ will affect the universality and $F_{best}$ cannot be the unit for some target operations, the optimization of the parameters can still correct part of the errors. The other two kinds of coherent errors, i.e., errors on $\phi_x$ and $\lambda_x$, can be corrected thoroughly. Second, comparing with (a) and (b), the difference between them is the target operation, they show that, when the target operation is closer to $\theta=0$, the resilience of the best fidelities to the coherent errors will become stronger. Third, the asymmetry of the original fidelity with the axis of $\delta=0$ is induced by $\phi_x$ and $\lambda_x$ (see the comparison between the solid blue lines and dashed orange lines).

Based on the above analysis for fidelity, the third kind of fidelity we want to study is the average fidelity, which is also another metric to evaluate universality. The average fidelity can be defined as 

\begin{equation}
F_{j}^{ave}=\frac{1}{V_{k}}\int_{V_{k}}F_{j}(\theta,\phi,\lambda)d\theta d\phi d\lambda,
\end{equation}
where $V_{k}$ is the volume we choose to calculate the average fidelity in the parameter space, and $j$ can be chosen as $ori$ or $best$. Then we take $V_k=V_{all}$ and give two examples. The first one is for the averaged original fidelity, when $\theta_{x}=\pi/2+\delta$, $\phi_{x}=0$, $\lambda_{x}=0$, the average fidelity is $F_{ori}^{ave}=1-\frac{4\sin^{2}(\delta/2)-3\sin^{4}(\delta/2)}{3}$. The second one is for the average best fidelity, we can prove that, for any case, there is $F_{best}^{ave}=1-\frac{2|\delta|-\sin(2|\delta|)}{3\pi}$. In all cases, there should be $F_{best}^{ave}\geq F_{ori}^{ave}$. The average fidelity is another metric for the decompositions, or we can treat it as a supplement for the definition of universality.  When $F_{best}^{ave}=1$, the universality is not destroyed. If there are two cases with different types of coherent errors, both shapes of the volume that cannot be covered by the decompositions are regular and continuous in one period for every direction, the universality and average fidelity will give the same information for the comparison of these two cases, just as the results shown in this paper. If the shape of the uncovered volume is irregular or not continuous in one period, the universality and average fidelity may give different results for the comparison of the two cases, but it will not appear for the case as we discussed in this work, it may appear in systems that are more complicated.

\emph{Conclusion}--We study the universality of the universal single-qubit gate decomposition scheme with coherent errors. We divide the coherent errors into three types depending on the parameters in the decomposition scheme. The original universality will be destroyed by one kind of coherent error, while it will stay unaffected with other kinds of coherent errors. We give a definition and analytical solutions for the universality with coherent errors, and we also provide the analytical solutions for three kinds of fidelities, i.e., original, best and average fidelity, and discuss their differences and significance. Depending on our analysis, when applying the universal decomposition scheme to generate an arbitrary single-qubit gate, this work gives a prediction for the original and optimized fidelities, and it also gives the method to mitigate the coherent errors and improve the gate fidelity. The research method used in this work can extend to the two-qubit gate decomposition schemes. Furthermore, it will bring different thoughts for the study of expressivity and quantum compilation theory in multi-qubit quantum systems to enhance the realistic fidelities for quantum computing.

\bigskip
We thank Dong E. Liu for helpful discussions. This work was supported by the Beijing Natural Science Foundation (Grant No. Z190012), the National Natural Science Foundation of China (Grants No. 11890704, No. 12004042, No. 12104055, No. 12104056),
and the Key-Area Research and Development Program of Guang Dong Province (Grant No. 2018B030326001).

%

\setcounter{equation}{0}
\setcounter{figure}{0}
\appendix

\renewcommand*{\thefigure}{S\arabic{figure}}
\renewcommand*{\theequation}{S\arabic{equation}}

\clearpage
\onecolumngrid
\tableofcontents

\clearpage 
\bigskip

\section{The matrix representations for universal single-qubit gates}

Depending on the discussions in the main text, the value ranges for this material are $\theta\in[0,\pi]$, $\phi\in[0,2\pi)$, $\lambda\in[0,2\pi)$.

\bigskip
\subsection{Ideal case}
When the X and Z rotations can be implemented ideally, the matrix representation of the universal decomposition for the arbitrary single-qubit gate is

\begin{eqnarray}
U(\theta,\phi,\lambda) & = & Z_{\phi-\pi/2}X_{\pi/2}Z_{\pi-\theta}X_{\pi/2}Z_{\lambda-\pi/2}\label{Udeco}\\
 & = & e^{i(-\lambda/2-\phi/2)}\left[\begin{array}{cc}
\cos(\theta/2) & -ie^{i\lambda}\sin(\theta/2)\\
-ie^{i\phi}\sin(\theta/2) & e^{i(\lambda+\phi)}\cos(\theta/2)
\end{array}\right],
\end{eqnarray}
where $ e^{i(-\lambda/2-\phi/2)}$ is a global phase and can be discarded when considering a single-qubit gate, then there is
\begin{eqnarray}
U(\theta,\phi,\lambda) & = \left[\begin{array}{cc}
\cos(\theta/2) & -ie^{i\lambda}\sin(\theta/2)\\
-ie^{i\phi}\sin(\theta/2) & e^{i(\lambda+\phi)}\cos(\theta/2)
\end{array}\right].
\label{U}
\end{eqnarray}

\subsection{The case with coherent errors}
When there are coherent errors on $X_{\pi/2}$ gate, we can replace $X_{\pi/2}$
gate by a unitary gate $U_{X_{\pi/2}}$, where $U_{X_{\pi/2}}=\left[\begin{array}{cc}
\cos(\theta_{x}/2) & -ie^{i\lambda_{x}}\sin(\theta_{x}/2)\\
-ie^{i\phi_{x}}\sin(\theta_{x}/2) & e^{i(\lambda_{x}+\phi_{x})}\cos(\theta_{x}/2)
\end{array}\right]$, and substituting it into equation \ref{Udeco} and replacing the operations of $X_{\pi/2}$, we can get
\bigskip
\begin{eqnarray}
\tilde{U} & = & Z_{\phi-\pi/2}U_{X_{\pi/2}}Z_{\pi-\theta}U_{X_{\pi/2}}Z_{\lambda-\pi/2}\label{oridecom}\nonumber\\
 & = & Z_{\phi-\pi/2}U_{X_{\pi/2}}Z_{\pi-\theta}\left[\begin{array}{cc}
\cos(\theta_{x}/2) & -ie^{i\lambda_{x}}\sin(\theta_{x}/2)\nonumber\\
-ie^{i\phi_{x}}\sin(\theta_{x}/2) & e^{i(\lambda_{x}+\phi_{x})}\cos(\theta_{x}/2)
\end{array}\right]\left[\begin{array}{cc}
e^{-i(\lambda/2-\pi/4)} & 0\nonumber\\
0 & e^{i(\lambda/2-\pi/4)}
\end{array}\right]\\
 & = & Z_{\phi-\pi/2}U_{X_{\pi/2}}\left[\begin{array}{cc}
e^{-i(\pi/2-\theta/2)} & 0\nonumber\\
0 & e^{i(\pi/2-\theta/2)}
\end{array}\right]\left[\begin{array}{cc}
e^{-i(\lambda/2-\pi/4)}\cos(\theta_{x}/2) & -ie^{i(\lambda_{x}+\lambda/2-\pi/4)}\sin(\theta_{x}/2)\nonumber\\
-ie^{i(\phi_{x}-\lambda/2+\pi/4)}\sin(\theta_{x}/2) & e^{i(\lambda_{x}+\phi_{x}+\lambda/2-\pi/4)}\cos(\theta_{x}/2)
\end{array}\right]\nonumber\\
 & = & e^{i(-\lambda/2-\phi/2)}\left[\begin{array}{c}
e^{i\theta/2}\cos^{2}(\theta_{x}/2)+e^{-i\theta/2}e^{i(\lambda_{x}+\phi_{x})}\sin^{2}(\theta_{x}/2)\nonumber\\
\left[e^{i(-\theta/2+\phi+\lambda_{x}+2\phi_{x})}-e^{i(\theta/2+\phi+\phi_{x})}\right]\frac{\sin\theta_{x}}{2}
\end{array}\right.\nonumber\\
 &  & \left.\begin{array}{c}
\left[e^{i(-\theta/2+\lambda+2\lambda_{x}+\phi_{x})}-e^{i(\theta/2+\lambda+\lambda_{x})}\right]\frac{\sin\theta_{x}}{2}\\
e^{i(\theta/2+\lambda+\phi+\lambda_{x}+\phi_{x})}\sin^{2}(\theta_{x}/2)+e^{i(-\theta/2+\lambda+\phi+2\lambda_{x}+2\phi_{x})}\cos^{2}(\theta_{x}/2)
\end{array}\right]
\end{eqnarray}

\bigskip

Without loss of generality, we discard the global
phase and take $\tilde{U}=\left[\tilde{U}^{(11)}, \tilde{U}^{(12)}; \tilde{U}^{(21)}, \tilde{U}^{(22)}\right]$,

\begin{eqnarray}
\tilde{U}^{(11)}&=&e^{i\theta/2}\cos^{2}(\theta_{x}/2)+e^{i(-\theta/2+\lambda_{x}+\phi_{x})}\sin^{2}(\theta_{x}/2)\\
\tilde{U}^{(12)}&=&\left[e^{i(-\theta/2+\lambda+2\lambda_{x}+\phi_{x})}-e^{i(\theta/2+\lambda+\lambda_{x})}\right]\frac{\sin\theta_{x}}{2}\\
\tilde{U}^{(21)}&=&\left[e^{i(-\theta/2+\phi+\lambda_{x}+2\phi_{x})}-e^{i(\theta/2+\phi+\phi_{x})}\right]\frac{\sin\theta_{x}}{2}\\
\tilde{U}^{(22)}&=&e^{i(\theta/2+\lambda+\phi+\lambda_{x}+\phi_{x})}\sin^{2}(\theta_{x}/2)+e^{i(-\theta/2+\lambda+\phi+2\lambda_{x}+2\phi_{x})}\cos^{2}(\theta_{x}/2)
\end{eqnarray}

\bigskip
\section{The analysis for coherent errors}

This section is for the formula derivations of the cases with different kinds or the combinations of coherent errors.

\bigskip
\subsection{Coherent error on $\theta_{x}$}

In this case, there are $\theta_{x}=\pi/2+\delta$, $\lambda_{x}=0$ and $\phi_{x}=0$, then there is

\begin{eqnarray}
\tilde{U}_{Case1} & = & \left[\begin{array}{cc}
e^{i\theta/2}\cos^{2}(\theta_{x}/2)+e^{-i\theta/2}\sin^{2}(\theta_{x}/2) & \frac{\sin\theta_{x}}{2}[e^{i(-\theta/2+\lambda)}-e^{i(\theta/2+\lambda)}]\\
\frac{\sin\theta_{x}}{2}[e^{i(-\theta/2+\phi)}-e^{i(\theta/2+\phi)}] & e^{i(\theta/2+\lambda+\phi)}\sin^{2}(\theta_{x}/2)+e^{i(-\theta/2+\lambda+\phi)}\cos^{2}(\theta_{x}/2)
\end{array}\right]\nonumber\\
 & = & \left[\begin{array}{cc}
\cos(\theta/2)+i\sin(\theta/2)\cos\theta_{x} & -ie^{i\lambda}\sin(\theta/2)\sin\theta_{x}\\
-ie^{i\phi}\sin(\theta/2)\sin\theta_{x} & e^{i(\lambda+\phi)}[\cos(\theta/2)-i\sin(\theta/2)\cos\theta_{x}]
\end{array}\right]\nonumber\\
 & = & e^{i\gamma_1}\left[\begin{array}{cc}
\sqrt{1-\sin^{2}(\theta/2)\sin^{2}\theta_{x}} & -ie^{i\lambda_{c1}}\sin(\theta/2)\sin\theta_{x}\\
-ie^{i\phi_{c1}}\sin(\theta/2)\sin\theta_{x} & e^{i(\lambda_{c1}+\phi_{c1})}\sqrt{1-\sin^{2}(\theta/2)\sin^{2}\theta_{x}}
\end{array}\right]
\end{eqnarray}

where $\tan\gamma_1=\tan(\theta/2)\cos\theta_{x}$,
$\lambda_{c1}=\lambda-\gamma_1$ and $\phi_{c1}=\phi-\gamma_1$.

\subsection{Coherent errors on $\phi_{x}$ and $\lambda_{x}$}

If there are coherent errors on $\phi_x$ and $\lambda_x$, there is
 
\begin{eqnarray}
\tilde{U}_{Case2} & = & \left[\begin{array}{c}
\frac{1}{2}[\cos(\theta/2)+\cos(-\theta/2+\lambda_{x}+\phi_{x})+i\sin(\theta/2)+i\sin(-\theta/2+\lambda_{x}+\phi_{x})]\\
\frac{1}{2}e^{i(\phi+\phi_{x})}[-\cos(\theta/2)+\cos(-\theta/2+\lambda_{x}+\phi_{x})-i\sin(\theta/2)+i\sin(-\theta/2+\lambda_{x}+\phi_{x})]
\end{array}\right.\nonumber\\
 &  & \left.\begin{array}{c}
\frac{1}{2}e^{i(\lambda+\lambda_{x})}[-\cos(\theta/2)+\cos(-\theta/2+\lambda_{x}+\phi_{x})-i\sin(\theta/2)+i\sin(-\theta/2+\lambda_{x}+\phi_{x})]\\
\frac{1}{2}e^{i(\lambda+\lambda_{x}+\phi+\phi_{x})}[\cos(\theta/2)+\cos(-\theta/2+\lambda_{x}+\phi_{x})+i\sin(\theta/2)+i\sin(-\theta/2+\lambda_{x}+\phi_{x})]
\end{array}\right]\nonumber\\
 & = & e^{i\gamma_{2}}\left[\begin{array}{cc}
\sqrt{\frac{1}{2}+\frac{1}{2}\cos(\theta-\lambda_{x}-\phi_{x})} & e^{i(\lambda+\lambda_{x}+\gamma'_{2}-\gamma_{2})}\sqrt{\frac{1}{2}-\frac{1}{2}\cos(\theta-\lambda_{x}-\phi_{x})}\\
e^{i(\phi+\phi_{x}+\gamma'_{2}-\gamma_{2})}\sqrt{\frac{1}{2}-\frac{1}{2}\cos(\theta-\lambda_{x}-\phi_{x})} & e^{i(\lambda+\lambda_{x}+\phi+\phi_{x})}\sqrt{\frac{1}{2}+\frac{1}{2}\cos(\theta-\lambda_{x}-\phi_{x})}
\end{array}\right],
\end{eqnarray}
with $\tan\gamma_{2}=\tan\frac{\lambda_{x}+\phi_{x}}{2}$, $\tan\gamma'_{2}=-\frac{1}{\tan(\frac{\lambda_{x}+\phi_{x}}{2})}$. From the formulas of $\tan\gamma_{2}$ and $\tan\gamma'_{2}$, we can get $\gamma'_{2}-\gamma_{2}=n\pi-\frac{\pi}{2}$ ($n$ is
an arbitrary integer), taking $n=0$, we can get

\begin{equation}
\tilde{U}_{Case2}=e^{i\gamma_{2}}\left[\begin{array}{cc}
\sqrt{\frac{1}{2}+\frac{1}{2}\cos(\theta-\lambda_{x}-\phi_{x})} & -ie^{i(\lambda+\lambda_{x})}\sqrt{\frac{1}{2}-\frac{1}{2}\cos(\theta-\lambda_{x}-\phi_{x})}\\
-ie^{i(\phi+\phi_{x})}\sqrt{\frac{1}{2}-\frac{1}{2}\cos(\theta-\lambda_{x}-\phi_{x})} & e^{i(\lambda+\lambda_{x}+\phi+\phi_{x})}\sqrt{\frac{1}{2}+\frac{1}{2}\cos(\theta-\lambda_{x}-\phi_{x})}
\end{array}\right].
\end{equation}

Defining $\theta_{c2}\equiv\theta-\lambda_{x}-\phi_{x}$, $\phi_{c2}\equiv\phi+\phi_{x}$
and $\lambda_{c2}\equiv\lambda+\lambda_{x}$, and by applying the
coordinate translations to make $\theta_{c2}\in[0,\pi]$, $\phi_{c2}\in[0,2\pi)$
and $\lambda_{c2}\in[0,2\pi)$, there is

\begin{eqnarray}
\tilde{U}_{Case2} & = & e^{i\gamma_{2}}\left[\begin{array}{cc}
\sqrt{\frac{1}{2}+\frac{1}{2}\cos\theta_{c2}} & -ie^{i\lambda_{c2}}\sqrt{\frac{1}{2}-\frac{1}{2}\cos\theta_{c2}}\\
-ie^{i\phi_{c2}}\sqrt{\frac{1}{2}-\frac{1}{2}\cos\theta_{c2}} & e^{i(\lambda_{c2}+\phi_{c2})}\sqrt{\frac{1}{2}+\frac{1}{2}\cos\theta_{bar}}
\end{array}\right]\nonumber\\
 & = & e^{i\gamma_{2}}\left[\begin{array}{cc}
\cos(\theta_{c2}/2) & -ie^{i\lambda_{c2}}\sin(\theta_{c2}/2)\\
-ie^{i\phi_{c2}}\sin(\theta_{c2}/2) & e^{i(\lambda_{c2}+\phi_{c2})}\cos(\theta_{c2}/2)
\end{array}\right],
\end{eqnarray}
which is equivalent to the equation (\ref{U}).

\subsection{Coherent errors on $\text{\ensuremath{\theta_{x}}}$, $\phi_{x}$ and $\lambda_{x}$}

If there are errors on $\ensuremath{\theta_{x}}$, $\phi_{x}$ and
$\lambda_{x}$, there is $\tilde{U}_{Case3}=\left[\tilde{U}_{Case3}^{(11)}, \tilde{U}_{Case3}^{(12)}; \tilde{U}_{Case3}^{(21)}, \tilde{U}_{Case3}^{(22)}\right]$, where 

\begin{eqnarray}
\tilde{U}_{Case3}^{(11)}&=&e^{i\theta/2}\cos^{2}(\theta_{x}/2)+e^{-i\theta/2}e^{i(\lambda_{x}+\phi_{x})}\sin^{2}(\theta_{x}/2)\\
\tilde{U}_{Case3}^{(12)}&=&\frac{\sin\theta_{x}}{2}[e^{i(-\theta/2+\lambda+2\lambda_{x}+\phi_{x})}-e^{i(\theta/2+\lambda+\lambda_{x})}]\\
\tilde{U}_{Case3}^{(21)}&=&\frac{\sin\theta_{x}}{2}[e^{i(-\theta/2+\phi+\lambda_{x}+2\phi_{x})}-e^{i(\theta/2+\phi+\phi_{x})}]\\
\tilde{U}_{Case3}^{(22)}&=&e^{i(\theta/2+\lambda+\phi+\lambda_{x}+\phi_{x})}\sin^{2}(\theta_{x}/2)+e^{i(-\theta/2+\lambda+\phi+2\lambda_{x}+2\phi_{x})}\cos^{2}(\theta_{x}/2)
\end{eqnarray}

For $\tilde{U}_{Case3}^{(11)}$, there is

\begin{eqnarray}
\tilde{U}_{Case3}^{(11)} & = & e^{i\theta/2}\cos^{2}(\theta_{x}/2)+e^{i(-\theta/2+\lambda_{x}+\phi_{x})}\sin^{2}(\theta_{x}/2)\nonumber\\
 & = & \cos(\theta/2)\cos^{2}(\theta_{x}/2)+\cos(-\theta/2+\lambda_{x}+\phi_{x})\sin^{2}(\theta_{x}/2)\nonumber\\
 &  & +i\sin(\theta/2)\cos^{2}(\theta_{x}/2)+i\sin(-\theta/2+\lambda_{x}+\phi_{x})\sin^{2}(\theta_{x}/2)\nonumber\\
 & = & e^{i\gamma_{3}}\sqrt{1-\sin^{2}\theta_{x}\sin^{2}(\frac{\theta-\lambda_{x}-\phi_{x}}{2})},
\end{eqnarray}
with $\tan\gamma_{3}=\frac{\sin(\theta/2)\cos^{2}(\theta_{x}/2)+\sin(-\theta/2+\phi_{x}+\lambda_{x})\sin^{2}(\theta_{x}/2)}{\cos(\theta/2)\cos^{2}(\theta_{x}/2)+\cos(-\theta/2+\phi_{x}+\lambda_{x})\sin^{2}(\theta_{x}/2)}$.

For $\tilde{U}_{Case3}^{(12)}$, there is

\begin{eqnarray}
\tilde{U}_{Case3}^{(12)} & = & \frac{\sin\theta_{x}}{2}[e^{i(-\theta/2+\lambda+2\lambda_{x}+\phi_{x})}-e^{i(\theta/2+\lambda+\lambda_{x})}]\nonumber\\
 & = & -\frac{\sin\theta_{x}}{2}e^{i(\lambda+\lambda_{x})}[\cos(\theta/2)-\cos(-\theta/2+\lambda_{x}+\phi_{x})+i\sin(\theta/2)-i\sin(-\theta/2+\lambda_{x}+\phi_{x})]\nonumber\\
 & = & -e^{i(\lambda+\lambda_{x}+\gamma_{4})}\sin\theta_{x}\sin|\frac{\theta-\lambda_{x}-\phi_{x}}{2}|,
\end{eqnarray}
with $\tan\gamma_{4}=\frac{\sin(\theta/2)-\sin(-\theta/2+\lambda_{x}+\phi_{x})}{\cos(\theta/2)-\cos(-\theta/2+\lambda_{x}+\phi_{x})}$.

For $\tilde{U}_{Case3}^{(21)}$, there is

\begin{eqnarray}
\tilde{U}_{Case3}^{(21)} & = & \frac{\sin\theta_{x}}{2}[e^{i(-\theta/2+\phi+\lambda_{x}+2\phi_{x})}-e^{i(\theta/2+\phi+\phi_{x})}]\nonumber\\
 & = & -\frac{\sin\theta_{x}}{2}e^{i(\phi+\phi_{x})}[\cos(\theta/2)-\cos(-\theta/2+\lambda_{x}+\phi_{x})+i\sin(\theta/2)-i\sin(-\theta/2+\lambda_{x}+\phi_{x})]\nonumber\\
 & = & -e^{i(\phi+\phi_{x}+\gamma_{4})}\sin\theta_{x}\sin|\frac{\theta-\lambda_{x}-\phi_{x}}{2}|.
\end{eqnarray}

For $\tilde{U}_{Case3}^{(22)}$, there is

\begin{eqnarray}
\tilde{U}_{Case3}^{(22)} & = & e^{i(\theta/2+\lambda+\phi+\lambda_{x}+\phi_{x})}\sin^{2}(\theta_{x}/2)+e^{i(-\theta/2+\lambda+\phi+2\lambda_{x}+2\phi_{x})}\cos^{2}(\theta_{x}/2)\nonumber\\
 & = & e^{i(\lambda+\phi+\lambda_{x}+\phi_{x})}[e^{i\theta/2}\sin^{2}(\theta_{x}/2)+e^{i(-\theta/2+\lambda_{x}+\phi_{x})}\cos^{2}(\theta_{x}/2)]\nonumber\\
 & = & e^{i(\lambda+\phi+\lambda_{x}+\phi_{x}+\gamma_{5})}\sqrt{1-\sin^{2}\theta_{x}\sin^{2}(\frac{\theta-\lambda_{x}-\phi_{x}}{2})},
\end{eqnarray}
with $\tan\gamma_{5}=\frac{\sin(\theta/2)\sin^{2}(\theta_{x}/2)+\sin(-\theta/2+\lambda_{x}+\phi_{x})\cos^{2}(\theta_{x}/2)}{\cos(\theta/2)\sin^{2}(\theta_{x}/2)+\cos(-\theta/2+\lambda_{x}+\phi_{x})\cos^{2}(\theta_{x}/2)}$,

The $\tilde{U}_{Case3}$ matrix can be written as

\begin{eqnarray}
\tilde{U}_{Case3} & = & \left[\begin{array}{cc}
e^{i\gamma_{3}}\sqrt{1-\sin^{2}\theta_{x}\sin^{2}(\frac{\theta-\lambda_{x}-\phi_{x}}{2})} & -ie^{i(\lambda+\lambda_{x}+\gamma_{4}-\frac{\pi}{2})}\sin\theta_{x}\sin|\frac{\theta-\lambda_{x}-\phi_{x}}{2}|\\
-ie^{i(\phi+\phi_{x}+\gamma_{4}-\frac{\pi}{2})}\sin\theta_{x}\sin|\frac{\theta-\lambda_{x}-\phi_{x}}{2}| & e^{i(\lambda+\phi+\phi_{x}+\lambda_{x}+\gamma_{5})}\sqrt{1-\sin^{2}\theta_{x}\sin^{2}(\frac{\theta-\lambda_{x}-\phi_{x}}{2})}
\end{array}\right]\nonumber\\
 & = & e^{i\gamma_{3}}\left[\begin{array}{cc}
\sqrt{1-\sin^{2}\theta_{x}\sin^{2}(\frac{\theta-\lambda_{x}-\phi_{x}}{2})} & -ie^{i(\lambda+\lambda_{x}+\gamma_{4}-\frac{\pi}{2}-\gamma_{3})}\sin\theta_{x}\sin|\frac{\theta-\lambda_{x}-\phi_{x}}{2}|\\
-ie^{i(\phi+\phi_{x}+\gamma_{4}-\frac{\pi}{2}-\gamma_{3})}\sin\theta_{x}\sin|\frac{\theta-\lambda_{x}-\phi_{x}}{2}| & e^{i(\lambda+\phi+\lambda_{x}+\phi_{x}+\gamma_{5}-\gamma_{3})}\sqrt{1-\sin^{2}\theta_{x}\sin^{2}(\frac{\theta-\lambda_{x}-\phi_{x}}{2})}
\end{array}\right].
\end{eqnarray}

Defining $\lambda_{c3}\equiv\lambda+\lambda_{x}+\gamma_{4}-\frac{\pi}{2}-\gamma_{3}$,
$\phi_{c3}\equiv\phi+\phi_{x}+\gamma_{4}-\frac{\pi}{2}-\gamma_{3}$,
there is $\lambda_{c3}+\phi_{c3}=\lambda+\phi+\lambda_{x}+\phi_{x}+2\gamma_{4}-2\gamma_{3}-\pi$.
We can calculate the relationship between $2\gamma_{4}-2\gamma_{3}-\pi$
and $\gamma_{5}-\gamma_{3}$, i.e., the relationship between $\gamma_{4}-\gamma_{3}$
and $\gamma_{5}-\gamma_{4}+\pi$.

\begin{eqnarray}
\tan(\gamma_{4}-\gamma_{3}) & = & \frac{\tan\gamma_{4}-\tan\gamma_{3}}{1+\tan\gamma_{4}\tan\gamma_{3}}\nonumber\\
 & = & \frac{\sin(\theta-\phi_{x}-\lambda_{x})}{\cos\theta_{x}-\cos(\theta-\lambda_{x}-\phi_{x})\cos\theta_{x}},
\end{eqnarray}

\begin{eqnarray}
\tan(\gamma_{5}-\gamma_{4}+\pi) & = & \tan(\gamma_{5}-\gamma_{4})\nonumber\\
 & = & \frac{\tan\gamma_{5}-\tan\gamma_{4}}{1+\tan\gamma_{5}\tan\gamma_{4}}\nonumber\\
 & = & \frac{\sin(\theta-\lambda_{x}-\phi_{x})}{\cos\theta_{x}-\cos(-\theta+\lambda_{x}+\phi_{x})\cos\theta_{x}},
\end{eqnarray}
there is $\tan(\gamma_{4}-\gamma_{3})=\tan(\gamma_{5}-\gamma_{4}+\pi)$,
let $2\gamma_{4}-2\gamma_{3}-\pi=\gamma_{5}-\gamma_{3}$. The $\tilde{U}_{Case3}$
matrix can be written as

\begin{eqnarray}
\tilde{U}_{Case3} & = & e^{i\gamma_{3}}\left[\begin{array}{cc}
\sqrt{1-\sin^{2}\theta_{x}\sin^{2}(\frac{\theta-\lambda_{x}-\phi_{x}}{2})} & -ie^{i\lambda_{c3}}\sin\theta_{x}\sin|\frac{\theta-\lambda_{x}-\phi_{x}}{2}|\\
-ie^{i\phi_{c3}}\sin\theta_{x}\sin|\frac{\theta-\lambda_{x}-\phi_{x}}{2}| & e^{i(\lambda_{c3}+\phi_{c3})}\sqrt{1-\sin^{2}\theta_{x}\sin^{2}(\frac{\theta-\lambda_{x}-\phi_{x}}{2})}
\end{array}\right].
\end{eqnarray}

From the above calculations, we know that

\[
|\tilde{U}_{Case3}^{(12)}|=|\tilde{U}_{Case3}^{(21)}|,
\]

\[
|\tilde{U}_{Case3}^{(11)}|=|\tilde{U}_{Case3}^{(22)}|,
\]

\[
|\tilde{U}_{Case3}^{(11)}|^{2}+|\tilde{U}_{Case3}^{(12)}|^{2}=1.
\]

Next is the discussion about the value range of $\tilde{U}_{Case3}^{(11)}$.

\begin{eqnarray}
|\tilde{U}_{Case3}^{(11)}| & = & \sqrt{1-\sin^{2}\theta_{x}\sin^{2}(\frac{\theta-\lambda_{x}-\phi_{x}}{2})}
\end{eqnarray}

The value range is $[|\cos\theta_{x}|,1]$, i.e., $[|\sin\delta|,1]$,
when $\theta_{x}=\pi/2$, i.e., $\delta=0$, the value range of $|\tilde{U}_{Case3}|$
is $[0,1]$.

\bigskip
\section{Fidelities}
Depending on the discussions of \emph{Case 1} to \emph{Case 3} in the main text, the value ranges for the coherent errors in this material are restricted to $\delta\in[-\pi/2,0)\cup(0,\pi/2]$ with $\theta_x=\pi/2+\delta$, $\phi_x\in(0,2\pi)$ and $\lambda_x\in(0,2\pi)$. 
This section gives the formula derivations of three kinds of fidelities.

\subsection{Original fidelity}

The original fidelity means that, we operate an single-qubit gate with the
non-ideal $X_{\pi/2}$ gate. We denote it as $F_{ori}$. Then there
is

\begin{eqnarray}
F_{ori} & = & \frac{Tr(U_{imp}^{\dagger}U_{imp})+|Tr(U_{tar}^{\dagger}U_{imp})|^{2}}{d(d+1)}\nonumber\\
 & = & \frac{2+|Tr(U_{tar}^{\dagger}U_{imp})|^{2}}{6}
\end{eqnarray}

\begin{eqnarray}
|Tr(U_{tar}^{\dagger}U_{imp})|^{2} & = & \left|Tr(\left[\begin{array}{cc}
\cos(\theta/2) & ie^{-i\phi}\sin(\theta/2)\nonumber\\
ie^{-i\lambda}\sin(\theta/2) & e^{-i(\lambda+\phi)}\cos(\theta/2)
\end{array}\right]\right.\nonumber\\
 &  & \left[\begin{array}{c}
e^{i\theta/2}\cos^{2}(\theta_{x}/2)+e^{i(-\theta/2+\lambda_{x}+\phi_{x})}\sin^{2}(\theta_{x}/2)\nonumber\\
\frac{\sin\theta_{x}}{2}[e^{i(-\theta/2+\phi+\lambda_{x}+2\phi_{x})}-e^{i(\theta/2+\phi+\phi_{x})}]
\end{array}\right.\nonumber\\
 &  & \left.\left.\begin{array}{c}
\frac{\sin\theta_{x}}{2}[e^{i(-\theta/2+\lambda+2\lambda_{x}+\phi_{x})}-e^{i(\theta/2+\lambda+\lambda_{x})}]\nonumber\\
e^{i(\theta/2+\lambda+\phi+\lambda_{x}+\phi_{x})}\sin^{2}(\theta_{x}/2)+e^{i(-\theta/2+\lambda+\phi+2\lambda_{x}+2\phi_{x})}\cos^{2}(\theta_{x}/2)
\end{array}\right])\right|^{2}\nonumber\\
 & = & |e^{i\theta/2}\cos(\theta/2)\cos^{2}(\theta_{x}/2)+e^{i(-\theta/2+\lambda_{x}+\phi_{x})}\cos(\theta/2)\sin^{2}(\theta_{x}/2)\nonumber\\
 &  & +i\sin(\theta/2)\frac{\sin\theta_{x}}{2}[e^{i(-\theta/2+\lambda_{x}+2\phi_{x})}-e^{i(\theta/2+\phi_{x})}]\nonumber\\
 &  & +i\sin(\theta/2)\frac{\sin\theta_{x}}{2}[e^{i(-\theta/2+2\lambda_{x}+\phi_{x})}-e^{i(\theta/2+\lambda_{x})}]\nonumber\\
 &  & +\cos(\theta/2)[e^{i(\theta/2+\lambda_{x}+\phi_{x})}\sin^{2}(\theta_{x}/2)+e^{i(-\theta/2+2\lambda_{x}+2\phi_{x})}\cos^{2}(\theta_{x}/2)]|^{2}\nonumber\\
 & = & |2\cos(\theta/2)\cos^{2}(\theta_{x}/2)\cos(\theta/2-\lambda_{x}-\phi_{x})e^{i(\lambda_{x}+\phi_{x})}\nonumber\\
 &  & +2\sin(\theta/2)\sin\theta_{x}\sin(\theta/2-\phi_{x}/2-\lambda_{x}/2)\cos(\phi_{x}/2-\lambda_{x}/2)e^{i(\lambda_{x}+\phi_{x})}\nonumber\\
 &  & +2\cos(\theta/2)\sin^{2}(\theta_{x}/2)\cos(\theta/2)e^{i(\lambda_{x}+\phi_{x})}|^{2}\nonumber\\
 & = & 4|\cos(\theta/2)\frac{1}{2}[\cos(\theta/2-\lambda_{x}-\phi_{x})+\cos(\theta/2)]\nonumber\\
 &  & +\cos(\theta/2)\cos\theta_{x}\frac{1}{2}[\cos(\theta/2-\lambda_{x}-\phi_{x})-\cos(\theta/2)]\nonumber\\
 &  & +\sin(\theta/2)\sin\theta_{x}\sin(\theta/2-\phi_{x}/2-\lambda_{x}/2)\cos(\phi_{x}/2-\lambda_{x}/2)|^{2}\nonumber\\
 & = & 4|\cos(\theta/2)\cos(\theta/2-\lambda_{x}/2-\phi_{x}/2)\cos(\lambda_{x}/2+\phi_{x}/2)\nonumber\\
 &  & +\cos(\theta/2)\sin(\theta/2-\lambda_{x}/2-\phi_{x}/2)\sin(\lambda_{x}/2+\phi_{x}/2)\cos\theta_{x}\nonumber\\
 &  & +\sin(\theta/2)\sin(\theta/2-\lambda_{x}/2-\phi_{x}/2)\cos(\phi_{x}/2-\lambda_{x}/2)\sin\theta_{x}|^{2}\label{fidall}
\end{eqnarray}

Analysis:

(1) When $\theta_{x}=\pi/2$, $\phi_{x}=0$, $\lambda_{x}=0$, there
is $|Tr(U_{tar}^{\dagger}U_{imp})|^{2}=4$ and $F_{ori}=1$, 

(2) When $\theta_{x}=\pi/2$, $\phi_{x}=0$, $\lambda_{x}\ne0$, there
is $|Tr(U_{tar}^{\dagger}U_{imp})|^{2}=4\cos^{4}(\lambda_{x}/2)$,

\begin{equation}
F_{ori}=\frac{1+2\cos^{4}(\lambda_{x}/2)}{3},
\label{fid2}
\end{equation}

(3) When $\theta_{x}=\pi/2$, $\phi_{x}\ne0$, $\lambda_{x}=0$, there
is $|Tr(U_{tar}^{\dagger}U_{imp})|^{2}=4\cos^{4}(\phi_{x}/2)$, 

\begin{equation}
F_{ori}=\frac{1+2\cos^{4}(\phi_{x}/2)}{3},
\label{fid3}
\end{equation}

(4) When $\theta_{x}=\pi/2$, $\phi_{x}\ne0$, $\lambda_{x}\ne0$,
there is $|Tr(U_{tar}^{\dagger}U_{imp})|^{2}=4|\cos(\frac{\lambda_{x}}{2}+\frac{\phi_{x}}{2})\cos\frac{\lambda_{x}}{2}\cos\frac{\phi_{x}}{2}-\cos(\theta-\frac{\lambda_{x}}{2}-\frac{\phi_{x}}{2})\sin\frac{\lambda_{x}}{2}\sin\frac{\phi_{x}}{2}|^{2}$,

\begin{equation}
F_{ori}=\frac{1+2|\cos(\frac{\lambda_{x}}{2}+\frac{\phi_{x}}{2})\cos\frac{\lambda_{x}}{2}\cos\frac{\phi_{x}}{2}-\cos(\theta-\frac{\lambda_{x}}{2}-\frac{\phi_{x}}{2})\sin\frac{\lambda_{x}}{2}\sin\frac{\phi_{x}}{2}|^{2}}{3},
\label{fid4}
\end{equation}

(5) When $\theta_{x}=\pi/2+\delta$, $\phi_{x}=0$, $\lambda_{x}=0$,
there is $|Tr(U_{tar}^{\dagger}U_{imp})|^{2}=4|\cos^{2}(\theta/2)+\sin^{2}(\theta/2)\cos\delta|^{2}$,

\begin{equation}
F_{ori}=\frac{1+2[1-2\sin^{2}(\delta/2)\sin^{2}(\theta/2)]^{2}}{3}.
\label{fid5}
\end{equation}

\subsection{Best fidelity}

The second kind of fidelity is the optimized one, which is named
best fidelity and denoted as $F_{best}$.

The intended single qubit gate is $\left[\begin{array}{cc}
\cos(\theta/2) & -ie^{i\lambda}\sin(\theta/2)\\
-ie^{i\phi}\sin(\theta/2) & e^{i(\lambda+\phi)}\cos(\theta/2)
\end{array}\right]$, the implemented single qubit gate is $\tilde{U}_{Case3}$, according
to the above analysis, we can take $\lambda_{c3}=\lambda$, $\phi_{c3}=\phi$,
and get the corrected single qubit gate:

\begin{equation}
U_{cor}=e^{i\gamma_{3}}\left[\begin{array}{cc}
\sqrt{1-\sin^{2}\theta_{x}\sin^{2}(\frac{\theta_{imp}-\lambda_{x}-\phi_{x}}{2})} & -ie^{i\lambda}\sin\theta_{x}\sin|\frac{\theta_{imp}-\lambda_{x}-\phi_{x}}{2}|\\
-ie^{i\phi}\sin\theta_{x}\sin|\frac{\theta_{imp}-\lambda_{x}-\phi_{x}}{2}| & e^{i(\lambda+\phi)}\sqrt{1-\sin^{2}\theta_{x}\sin^{2}(\frac{\theta_{imp}-\lambda_{x}-\phi_{x}}{2})}
\end{array}\right]
\end{equation}

There is:

\begin{eqnarray}
|Tr(U^{\dagger}U_{cor})|^{2} & = & |\left[\begin{array}{cc}
\cos(\theta/2) & ie^{-i\phi}\sin(\theta/2)\\
ie^{-i\lambda}\sin(\theta/2) & e^{-i(\lambda+\phi)}\cos(\theta/2)
\end{array}\right]\nonumber\\
 &  & \left[\begin{array}{cc}
\sqrt{1-\sin^{2}\theta_{x}\sin^{2}(\frac{\theta_{imp}-\lambda_{x}-\phi_{x}}{2})} & -ie^{i\lambda}\sin\theta_{x}\sin|\frac{\theta_{imp}-\lambda_{x}-\phi_{x}}{2}|\\
-ie^{i\phi}\sin\theta_{x}\sin|\frac{\theta_{imp}-\lambda_{x}-\phi_{x}}{2}| & e^{i(\lambda+\phi)}\sqrt{1-\sin^{2}\theta_{x}\sin^{2}(\frac{\theta_{imp}-\lambda_{x}-\phi_{x}}{2})}
\end{array}\right]|^{2}\nonumber\\
 & = & 4|\cos(\theta/2)\sqrt{1-\sin^{2}\theta_{x}\sin^{2}(\frac{\theta_{imp}-\lambda_{x}-\phi_{x}}{2})}+\sin(\theta/2)\sin\theta_{x}\sqrt{\sin^{2}\frac{\theta_{imp}-\lambda_{x}-\phi_{x}}{2}}|^{2}
\end{eqnarray}

Defining $f(a)=\cos(\theta/2)\sqrt{1-\sin^{2}\theta_{x}a}+\sin(\theta/2)\sin\theta_{x}\sqrt{a}$,
where $a=\sin^{2}(\frac{\theta_{imp}-\phi_{x}-\lambda_{x}}{2})$ and
$a\in[0,1]$, there is

\begin{equation}
\partial f(a)/\partial a=\cos(\theta/2)\frac{-\sin^{2}\theta_{x}}{2\sqrt{1-\sin^{2}\theta_{x}a}}+\sin(\theta/2)\sin\theta_{x}\frac{1}{2\sqrt{a}}
\end{equation}

Let $\partial f(a)/\partial a|_{a=a_{0}}=0$, we can get,

\begin{eqnarray}
\cos(\theta/2)\frac{\sin^{2}\theta_{x}}{2\sqrt{1-\sin^{2}\theta_{x}a_{0}}} & = & \sin(\theta/2)\sin\theta_{x}\frac{1}{2\sqrt{a_{0}}}\nonumber\\
\frac{\cos(\theta/2)\sin^{2}\theta_{x}}{\sin(\theta/2)\sin\theta_{x}} & = & \frac{\sqrt{1-\sin^{2}\theta_{x}a_{0}}}{\sqrt{a_{0}}}\nonumber\\
a_{0}\cos^{2}(\theta/2)\sin^{2}\theta_{x} & = & \sin^{2}(\theta/2)-a_{0}\sin^{2}(\theta/2)\sin^{2}\theta_{x}\nonumber\\
a_{0} & = & \frac{\sin^{2}(\theta/2)}{\sin^{2}\theta_{x}}
\end{eqnarray}

The value of $\partial^{2}f(a)/\partial a^{2}$ at the $a_{0}$ point and with $\theta\in[0,\pi]$
is,

\begin{eqnarray}
\partial^{2}f(a)/\partial a^{2}|_{a=a_{0}} & = & \frac{-\cos(\theta/2)\sin^{4}\theta_{x}}{4(1-\sin^{2}\theta_{x}a)\sqrt{1-\sin^{2}\theta_{x}a}}|_{a=a_{0}}-\frac{1}{4}\sin(\theta/2)\sin\theta_{x}\frac{1}{a\sqrt{a}}|_{a=a_{0}}\nonumber\\
 & = & -\frac{1}{4}\frac{\sin^{4}\theta_{x}}{\cos^{2}(\theta/2)}-\frac{1}{4}\frac{\sin^{4}\theta_{x}}{\sin^{2}(\theta/2)}\nonumber\\
 & = & -\frac{\sin^{4}\theta_{x}}{4\cos^{2}(\theta/2)\sin^{2}(\theta/2)}<0
\end{eqnarray}

Since there is $f(a)>0$, $f(a)$
has the maximum value when $a$ takes the value of $a_{0}$, and from $a=0$ to $a=a_{0}$, the value of
$f(a)$ increases. If $a_{0}\le1$, $a$ can be equal to $a_{0}$,
and the best fidelity is at $a=a_{0}$, the best fidelity is

\begin{eqnarray}
F_{best} & = & \frac{2+4|\cos^{2}(\theta/2)+\sin^{2}(\theta/2)|^{2}}{6}\nonumber\\
 & = & 1
\end{eqnarray}

If $a_{0}>1$, i.e., $\sin^{2}(\theta/2)>\sin^{2}\theta_{x}$, the
best fidelity is achieved at $a=1$, if $\cos\theta_{x}>0$, there is $\text{\ensuremath{\theta_{x}=\frac{\pi}{2}+\delta}}$,
$\delta\in[-\frac{\pi}{2},0)$ and 

\begin{eqnarray}
F_{best} & = & \frac{2+4|\cos(\theta/2)\cos\theta_{x}+\sin(\theta/2)\sin\theta_{x}|^{2}}{6}\nonumber\\
 & = & \frac{1+2\cos^{2}(\theta/2-\theta_{x})}{3}\nonumber\\
 & = & \frac{1+2\sin^{2}(\theta/2+|\delta|)}{3}
\label{fbest1}
\end{eqnarray}

As there is $\sin^{2}(\theta/2)>\sin^{2}\theta_{x}$, if $\theta/2-\theta_{x}=n\pi$
($n$ is an interge), there is $\sin^{2}(\theta/2)=\sin^{2}\theta_{x}$,
it conflicts with $\sin^{2}(\theta/2)>\sin^{2}\theta_{x}$, then
$\cos^{2}(\theta/2-\theta_{x})\ne1$ and $F_{best}=\frac{1+2\sin^{2}(\theta/2+|\delta|)}{3}<1$.

If $\cos\theta_{x}<0$, there is $\text{\ensuremath{\theta_{x}=\frac{\pi}{2}+\delta}}$,
$\delta\in(0,\frac{\pi}{2}]$ and

\begin{eqnarray}
F_{best} & = & \frac{2+4|-\cos(\theta/2)\cos\theta_{x}+\sin(\theta/2)\sin\theta_{x}|^{2}}{6}\nonumber\\
 & = & \frac{1+2\cos^{2}(\theta/2+\theta_{x})}{3}\nonumber\\
 & = & \frac{1+2\sin^{2}(\theta/2+|\delta|)}{3}
\label{fbest2}
\end{eqnarray}

As there is $\sin^{2}(\theta/2)>\sin^{2}\theta_{x}$, if $\theta/2+\theta_{x}=n\pi$ ($n$ is an interge), there is $\sin^{2}(\theta/2)=\sin^{2}\theta_{x}$,
it conflicts with $\sin^{2}(\theta/2)>\sin^{2}\theta_{x}$, so that
$\cos^{2}(\theta/2+\theta_{x})\ne1$ and $F_{best}=\frac{1+2\sin^{2}(\theta/2+|\delta|)}{3}<1$.

Note that, this formula of $F_{best}$ is for $\theta\in[0,\pi]$, and from the derivation process, we know that, for $\theta\in[-\pi, 0)$, there should be a minus sign before $\theta$ in $F_{best}$.

\subsection{Average fidelity}

The average fidelity can be defined as 

\begin{equation}
F_{j}^{ave}=\frac{1}{V_{k}}\int_{V_{k}}F_{j}(\theta,\phi,\lambda)d\theta d\phi d\lambda
\end{equation}

The volume $V_{k}$ is the volume we choose to calculate the fidelity,
and $j$ can be chosen as $ori$ or $best$.

The average original fidelity can be calculated
as,

\begin{equation}
F_{ori}^{ave}=\frac{1}{V_{all}}\int_{V_{all}}F_{ori}(\theta,\phi,\lambda)d\theta d\phi d\lambda
\end{equation}

For the special cases, we have,

(1) When $\theta_{x}=\pi/2$, $\phi_{x}=0$, $\lambda_{x}=0$, there
is $|Tr(U_{tar}^{\dagger}U_{imp})|^{2}=4$ and $F_{ori}=1$ for all the
$(\theta,\phi,\lambda)$, then there is $F_{ori}^{ave}=1$; 

(2) When $\theta_{x}=\pi/2$, $\phi_{x}=0$, $\lambda_{x}\ne0$, there
is $|Tr(U_{tar}^{\dagger}U_{imp})|^{2}=4\cos^{4}(\lambda_{x}/2)$, $F_{ori}=\frac{2+4\cos^{4}(\lambda_{x}/2)}{6}$,
which only depends on the value of $\lambda_{x}$, then there is $F_{ori}^{ave}=\frac{1+2\cos^{4}(\lambda_{x}/2)}{3}$;

(3) When $\theta_{x}=\pi/2$, $\phi_{x}\ne0$, $\lambda_{x}=0$, there
is $|Tr(U_{tar}^{\dagger}U_{imp})|^{2}=4\cos^{4}(\phi_{x}/2)$, $F_{ori}=\frac{2+4\cos^{4}(\phi_{x}/2)}{6}$,
which only depends on the value of $\phi_{x}$, then there is $F_{ori}^{ave}=\frac{1+2\cos^{4}(\phi_{x}/2)}{3}$;

(4) When $\theta_{x}=\pi/2$, $\phi_{x}\ne0$, $\lambda_{x}\ne0$,
there is $|Tr(U_{tar}^{\dagger}U_{imp})|^{2}=4|\cos(\frac{\lambda_{x}}{2}+\frac{\phi_{x}}{2})\cos\frac{\lambda_{x}}{2}\cos\frac{\phi_{x}}{2}-\cos(\theta-\frac{\lambda_{x}}{2}-\frac{\phi_{x}}{2})\sin\frac{\lambda_{x}}{2}\sin\frac{\phi_{x}}{2}|^{2}$,
then there is

\begin{eqnarray}
F_{ori}^{ave} & = & \frac{1}{\pi}\int_{0}^{\pi}\frac{2+4|\cos(\frac{\lambda_{x}+\phi_{x}}{2})\cos\frac{\lambda_{x}}{2}\cos\frac{\phi_{x}}{2}-\cos(\theta-\frac{\lambda_{x}+\phi_{x}}{2})\sin\frac{\lambda_{x}}{2}\sin\frac{\phi_{x}}{2}|^{2}}{6}d\theta\nonumber\\
 & = & \frac{1}{3}+\frac{2\cos^{2}(\frac{\lambda_{x}+\phi_{x}}{2})\cos^{2}\frac{\lambda_{x}}{2}\cos^{2}\frac{\phi_{x}}{2}}{3}\nonumber\\
 &  & +\frac{\sin^{2}\frac{\lambda_{x}}{2}\sin^{2}\frac{\phi_{x}}{2}}{3}\nonumber\\
 &  & -\frac{\sin\lambda_{x}\sin\phi_{x}\sin(\lambda_{x}+\phi_{x})}{3\pi}
\end{eqnarray}

(5) When $\theta_{x}=\pi/2+\delta$, $\phi_{x}=0$, $\lambda_{x}=0$,
there is $F_{ori}=\frac{1+2[1-2\sin^{2}(\delta/2)\sin^{2}(\theta/2)]^{2}}{3}$.
We can calculate the average fidelity as,

\begin{eqnarray}
F_{ori}^{ave} & = & \frac{4\pi^{2}}{4\pi^{3}}\int_{0}^{\pi}\frac{1+2[1-2\sin^{2}(\delta/2)\sin^{2}(\theta/2)]^{2}}{3}d\theta\nonumber\\
 & = & 1+\frac{1}{\pi}\int_{0}^{\pi}\frac{-4\sin^{2}(\delta/2)+3\sin^{4}(\delta/2)}{3}d\theta\nonumber\\
 & = & 1-\frac{4\sin^{2}(\delta/2)-3\sin^{4}(\delta/2)}{3}
\end{eqnarray}

For the averaged best fidelity of any case, there is

\begin{eqnarray}
F_{best}^{ave} & = & \frac{4\pi^{2}}{4\pi^{3}}\int_{0}^{\pi}F_{best}(\theta)d\theta\nonumber\\
 & = & \frac{1}{\pi}[\int_{\pi-2|\delta|}^{\pi}\frac{1+2\sin^{2}(\theta/2+|\delta|)}{3}d\theta+\int_{0}^{\pi-2|\delta|}1d\theta\text{]}\nonumber\\
 & = & 1-\frac{2|\delta|-\sin(2|\delta|)}{3\pi}
\end{eqnarray}

\bigskip

\section{Another view}

\begin{figure}[h!]
\centering{\includegraphics[width=50mm]{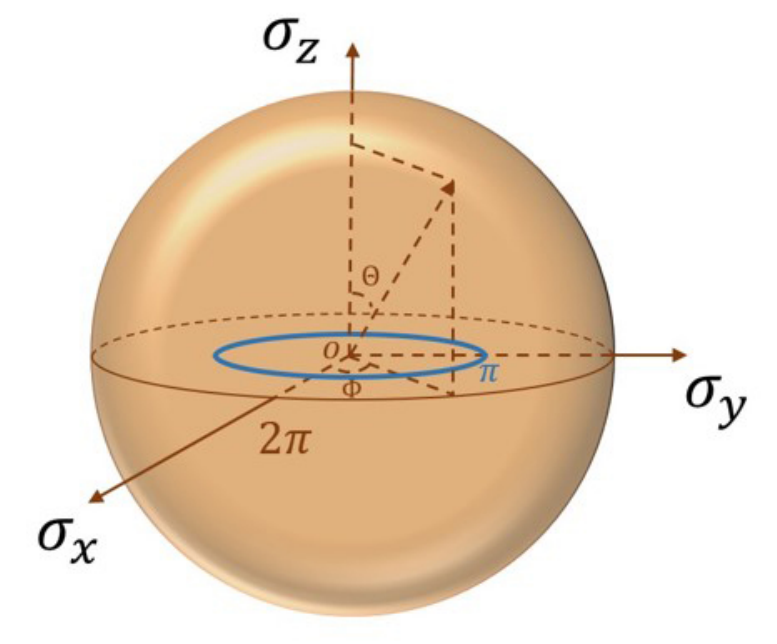}}
\caption{The diagram of the spherical coordinate, which can represent arbitrary coherent single-qubit operation. The vector in the spherical coordinate represents the rotation axis, and the length of the radius represents the rotation angle. The blue areas represent the parameter space, which cannot be covered by the single-qubit gate decomposition scheme with coherent errors.} 
\label{fball}
\end{figure}

There is an alternative perspective, which is that we can represent a single-qubit gate by a rotation in a spherical coordinate system. We define a vector $\vec{n}=(\sin\Theta\cos\Phi,\sin\Theta\sin\Phi,\cos\Theta)$ to represent a unit direction in a spherical coordinate with $\Theta$ being the polar angle and $\Phi$ being the azimuthal angle, $\omega$ is the length of the radius, as shown in figure \ref{fball}. We can write the SU(2) matrix in the spherical coordinate as $U_s(\hat{n},\omega) = I\cos(\omega/2)-i(\vec{\sigma}\cdot\hat{n})\sin(\omega/2)$, where $\vec{\sigma}=\Sigma_{l=1}^{3}\vec{e}_{l}\sigma_l$, $\sigma_l$ are three pauli matrices, $e_l$ is the unit vector in the $l$th direction.  $U_s(\hat{n},\omega)$ can represent the rotation of angle $\omega$ along the $\hat{n}$ direction. The matrix representation is $U_s(\hat{n},\omega)  =  e^{-i\Gamma}\left[U_s^{(11)}, U_s^{(12)}; U_s^{(21)}, U_s^{(22)}\right]$,
where $\tan\Gamma=\tan(\omega/2)\cos\Theta$ and
\begin{eqnarray}
U_s^{(11)}&=&\sqrt{1-\sin^{2}(\omega/2)\sin^{2}\Theta}, \\
U_s^{(12)}&= &-ie^{i(-\Phi+\Gamma)}\sin(\omega/2)\sin\Theta, \\
U_s^{(21)}&=&-ie^{i(\Phi+\Gamma)}\sin(\omega/2)\sin\Theta, \\
U_s^{(22)}&=&e^{2i\Gamma}U_s^{(11)}.
\end{eqnarray}

Compared with $\tilde{U}_{Case3}$, we can see that, if $|U_s^{(11)}|\in[|\sin\delta|,1]$ and $\delta\ne0$, the operations which satisfy $|U_s^{(11)}|\in[0,|\sin\delta|)$ cannot be realized. Let $|U_s^{(11)}|=0$, we can get $\omega=\pi$ and $\Theta=\pi/2$, this operation represents the rotation of angle $\pi$ along the axis in the $\sigma_{x}\sigma_{y}$ plan, we denote the set including such operations as $O_{\pi}^{xy}$, which is shown in figure \ref{fball} with the blue circle. When $\delta\ne0$, the operations of $O_{\pi}^{xy}$ and other operations near $O_{\pi}^{xy}$ cannot be covered by the decomposition scheme, which means that, the blue circle in figure \ref{fball} has a volume, and the size of the volume depends on the value of $|\delta|$, and the operations reside in this volume cannot be implemented with unit fidelity even after the parameter optimizations. If $|\delta|$
is small, e.g. $|\delta|\ll\pi$, $|U_s^{(11)}|=\sqrt{1-\frac{1}{2}\sin^{2}\Theta}$
can be realized with arbitrary value of $\Theta$, we can combine two rotations with $\omega=\pi/2$ to one rotation with $\omega=\pi$. However, it will increase the circuit length, whether or not to taking this way depends on the tradeoff between coherent and incoherent errors.

\bigskip
\section{Numerical verification}

\begin{figure}[h!]
\centering{\includegraphics[width=170mm]{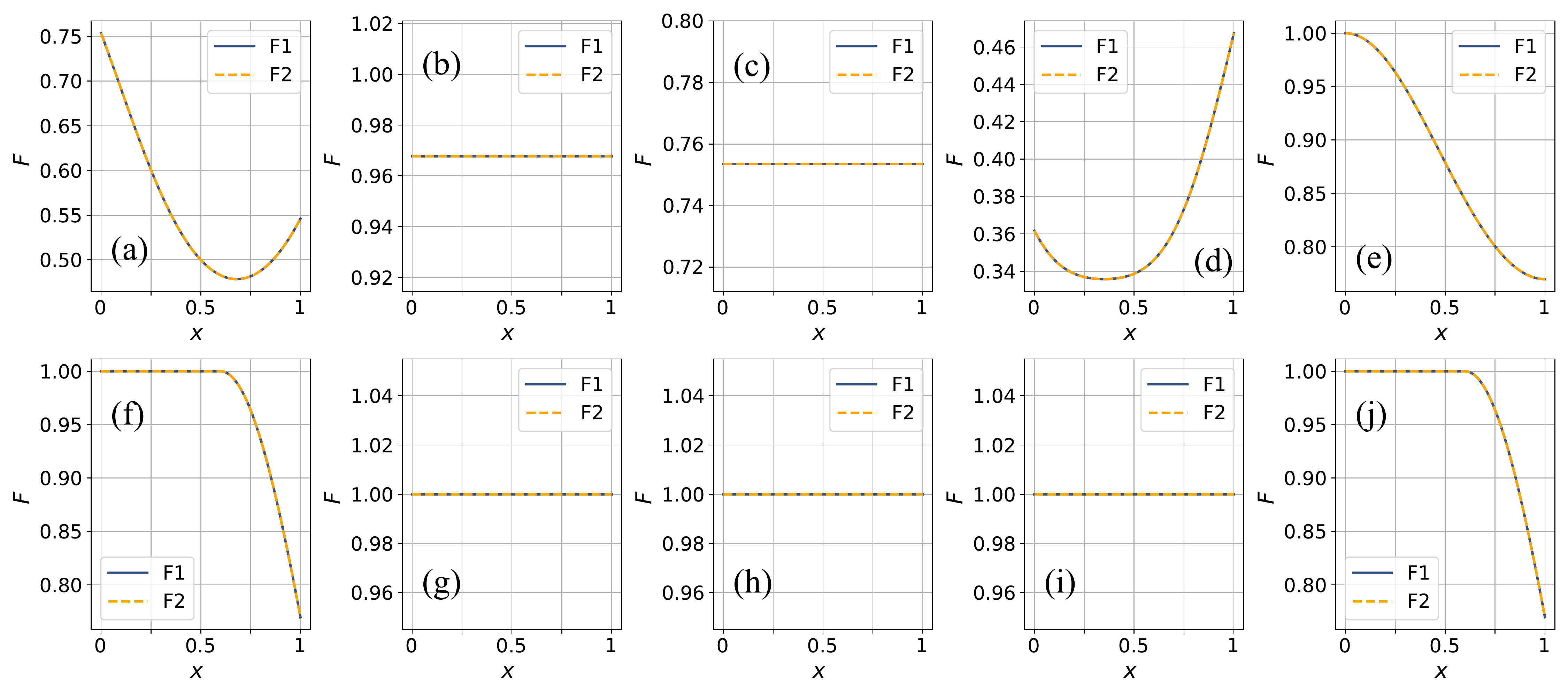}}
\caption{Numerical demonstrations for the theoretical results. The values of the parameters for the target single-qubit gates are chosen as $\theta/\pi=x$, $\phi/2\pi=x$ and $\lambda/2\pi=x$ for (a)-(i). For (j), we set $\theta/\pi=-x$, $\phi/1\pi=x$ and $\lambda/2\pi=x$. The values of the parameters for coherent errors are chosen as (a) $\theta_x/\pi=1/2+0.2$, $\phi_x/\pi=0.3$, $\lambda_x/\pi=0.2$. (b) $\theta_x/\pi=1/2$, $\phi_x=0$, $\lambda_x/\pi=0.1$. (c) $\theta_x/\pi=1/2$, $\phi_x/\pi=0.3$, $\lambda_x=0$. (d) $\theta_x/\pi=1/2$, $\phi_x/\pi=0.3$, $\lambda_x/\pi=0.4$. (e) $\theta_x/\pi=1/2+0.2$, $\phi_x=0$, $\lambda_x=0$, and the values for the parameters $\{\theta_x,\phi_x,\lambda_x\}$ of (f)-(j) are the same with (a)-(e) respectively. } 
\label{nd}
\end{figure}

To verify the theoretical results which we have got in the manuscript, we do some numerical simulations as shown in figure \ref{nd}. Figures \ref{nd}(a)-(e) are the demonstrations for the original fidelity, where $\rm F1$ is calculated by substituting the values of parameters into the matrix $U_{tar}$ (equation \ref{U}) and $U_{imp}$ (the original decomposition of \ref{oridecom}) directly and then obtaining the fidelity, and $\rm F2$ is got from the simplified theoretical formula, for example, the results of $\rm F2$ in (a) is got by using the result of \ref{fidall}, and which in (b)-(e) are got by using the results of \ref{fid2}, \ref{fid3}, \ref{fid4} and \ref{fid5}. Figures \ref{nd}(f)-(j) are the demonstrations for the best fidelity, where $\rm F1$ is the optimized results, which means that, for getting these results, we optimized the implemented parameters of $\{\theta_{imp}, \phi_{imp}, \lambda_{imp}\}$ to find the optimized fidelity, and $\rm F2$ is the theoretical results, which is got by using the formula of \ref{fbest2} for (f)-(i), and for (j), we set $\theta\in[-\pi,0]$, and $F_{best} = \frac{1+2\sin^{2}(-\theta/2+|\delta|)}{3}$, and then get the results of $\rm F2$. The theoretical results agree well with the numerical results.

\end{document}